\documentclass[aps,prb,10pt,amsmath,amssymb,twocolumn,longbibliography,superscriptaddress]{revtex4-2}
\usepackage{float}
\usepackage[pdftex]{graphicx,xcolor}
\usepackage{dcolumn}
\usepackage{bm}
\usepackage{ascmac}
\usepackage{array}
\usepackage{siunitx}
\usepackage{physics}
\usepackage{color}
\usepackage{soul}
\usepackage{hyperref}
\hypersetup{
     colorlinks=true,
     citecolor=blue,
     linkcolor=blue,
     }
\begin{document}

\title {Local expression of fractional corner charges in obstructed atomic insulators and relationship with the fractional disclination charges}
\author {Hidetoshi Wada}
\affiliation{
Department of Physics, Institute of Science Tokyo, 2-12-1 Ookayama, Meguro-ku, Tokyo 152-8551, Japan}

\author {Shuichi Murakami}
\affiliation{
Department of Physics, Institute of Science Tokyo, 2-12-1 Ookayama, Meguro-ku, Tokyo 152-8551, Japan}

\affiliation{
Department of Applied Physics, University of Tokyo, 7-3-1 Hongo, Bunkyo-ku, Tokyo 113-8656, Japan \\
}

\affiliation{
International Institute for Sustainability with Knotted Chiral Meta Matter (WPI-SKCM$^{2}$), Hiroshima University, Higashi-hiroshima, Hiroshima 739-0046, Japan}

\affiliation{
Center for Emergent Matter Science, RIKEN, Hirosawa, Wako, Saitama 351-0198, Japan}

\date{\today}

\begin{abstract}
In obstructed atomic insulators, fractionally quantized charges appear at the corners of the crystals in the shapes of vertex-transitive polyhedra, and are given by the filling anomaly divided by the number of corners. 
Recent studies reveal that the filling anomaly for the cases with genus $0$ is universally given by the total charge at the Wyckoff position $1a$. 
In this study, we rewrite the formula in terms of the degree of sharpness of the corner, and show that the corner charge formula also holds for cases with arbitrary genus. 
We also extend our formula to vertex-transitive shell polyhedra, which are closed or open polyhedra without the bulk region, with all the vertices related by symmetry.
Then, we show that the corner charges of such shell polyhedra are equal to the two-dimensional disclination charges of the corresponding disclinations. 
By identifying it with the disclination charge under the Wen-Zee action, we show that the coupling constant of the Wen-Zee action for a crystalline insulator is given by the total charge at the Wyckoff position at the disclination core.
\end{abstract}

\maketitle

\section{Introduction}
The study of topological phases of matter is one of the most important topics today \cite{PhysRevB.76.045302,PhysRevB.27.6083,PhysRevLett.98.106803,PhysRevLett.95.146802,PhysRevLett.95.226801,PhysRevB.74.195312,doi:10.1126/science.1148047,doi:10.1126/science.1133734,PhysRevB.78.045426,PhysRevLett.106.106802,PhysRevB.96.245115,PhysRevB.91.161105,PhysRevB.95.081107,PhysRevB.90.165114,PhysRevX.7.041069,PhysRevB.98.081110,PhysRevLett.119.246402,Schindler_2018,doi:10.1126/sciadv.aat0346,PhysRevResearch.2.012067,PhysRevLett.124.036803,PhysRevResearch.2.043131,PhysRevX.11.041064,PhysRevB.105.045126,PhysRevB.109.085114,PhysRevB.111.155305,PhysRevB.99.245151,PhysRevB.103.205123,PhysRevResearch.1.033074,PhysRevB.102.165120,Po_2017,SI_SA_Watanabe,Bradlyn_2017,PhysRevB.97.035139,Aroyo:xo5013,PhysRevB.112.094204}. 
In the topologically nontrivial phases of matter, there are gapless excitations localized at their boundaries, which are robust against continuous deformations, and they are characterized by the classification indices called topological invariants.  
Since the proposal of Chern insulators \cite{PhysRevB.76.045302,PhysRevB.27.6083} and $\mathbb{Z}_{2}$ topological insulators \cite{PhysRevLett.98.106803,PhysRevLett.95.146802,PhysRevLett.95.226801,PhysRevB.74.195312,doi:10.1126/science.1148047,doi:10.1126/science.1133734}, varieties of topological insulators have been discovered, such as topological crystalline insulators (TCIs) \cite{PhysRevLett.106.106802,PhysRevB.96.245115,PhysRevB.91.161105,PhysRevB.95.081107,PhysRevB.90.165114,PhysRevX.7.041069,PhysRevB.112.094204} protected by crystallographic symmetries and higher-order topological insulators (HOTIs) \cite{PhysRevB.98.081110,PhysRevLett.119.246402,Schindler_2018,doi:10.1126/sciadv.aat0346,PhysRevResearch.2.012067,PhysRevLett.124.036803,PhysRevResearch.2.043131} with gapless states on their hinges or corners. 
Furthermore, topological quantum chemistry (TQC) \cite{Bradlyn_2017,PhysRevB.97.035139,Aroyo:xo5013, MTQC} and the theory of symmetry-based indicators \cite{Po_2017, SI_SA_Watanabe} have allowed us to identify many topological materials \cite{catalogue0,catalogue1,catalogue2,catalogue3,catalogue4,catalogue6}. 

On the other hand, topologically trivial insulators, called atomic insulators (AIs), have been believed to have no topologically interesting features because their energy spectra are completely gapped including their boundaries. 
However, recent studies \cite{PhysRevB.105.045126,PhysRevB.109.085114,PhysRevB.99.245151,PhysRevB.111.155305,PhysRevX.11.041064,PhysRevB.103.205123,PhysRevResearch.1.033074,PhysRevB.102.165120} have revealed that some AIs called obstructed atomic insulators (OAIs) have fractionally quantized charges on their corners, which are characterized by a topological invariant called a filling anomaly \cite{PhysRevB.99.245151,PhysRevB.103.165109}.
The filling anomaly is given by the total charge of a finite-sized system preserving crystallographic symmetries which connect all corners of the system, and the fractional corner charges can be obtained by equivalent distribution of the filling anomaly to the corners.  
The filling anomaly is well understood in two dimensions, and is given by the total charge at the Wyckoff position (WP) $1a$ located at the center of the systems \cite{PhysRevB.103.205123,PhysRevB.102.165120}. 

In previous works \cite{PhysRevB.109.085114,PhysRevB.111.155305}, we constructed a complete list of corner charge formulas for real and $k$ spaces in three-dimensional crystals. 
In this case, the crystal shapes need to be the vertex-transitive polyhedra with genus $0$ \cite{polyhedra,bridges2002:320,ROBERTSON197079,https://doi.org/10.1112/jlms/s2-2.1.125,IsogonalPrismatoids}, and for all the corresponding crystal shapes for all space groups, we calculated the filling anomalies and corner charge formulas. 
As a result, the real-space formulas of the filling anomaly are universally given by the total charge at the bulk WP $1a$ located at the center of the crystal shapes, and the corner charge formulas are given by the filling anomaly divided by the number of corners, similar to the two-dimensional cases \cite{PhysRevB.103.205123}. 

Microscopically, we expect that the corner charge should be determined locally, i.e., it depends only on the shape of the corner.
Thus we address the question of why the corner charge depends on the number of corners, which does not seem to be a local property. 
In this study, we derive a local expression of the corner charge formulas, and thereby, construct general corner charge formulas in crystals with the forms of the vertex-transitive polyhedra with arbitrary genus. 
Then we introduce the concept of vertex-transitive shell polyhedra (VTSPs), which are closed or open polyhedra without the bulk region, with all the vertices related by symmetry. 
Moreover, we construct the relationship between the two-dimensional disclination charges and the corner charges in the corresponding VTSPs, which leads to a general formulation of the disclination charges in two dimensions. 
Finally, we show that the coupling constant of the Wen-Zee action \cite{PhysRevLett.69.953} in such crystalline insulators is given by the total charge at the WP, where the disclination core is located. 
We note that near the surfaces of the crystals, the electronic states and nuclear positions may shift slightly from the ideal configuration of crystals. 
Nevertheless, the fractional corner charge survives such deviation from perfect periodicity, as long as such deviation does not break the focused point-group symmetries, and the lattice periodicity along the hinges and the surfaces.
Thus, the fractional corner charge is robust against those surface effects, where the lattice periodicity in the bulk is preserved along the hinges and the surfaces.

This paper is organized as follows. 
In Sec.~\ref{Sec.2}, we briefly review the previous studies of the fractionally quantized corner charges. 
In Sec.~\ref{Sec.3}, we construct the local expression of the corner charge formulas. 
In Sec.~\ref{Sec.4}, we construct the corner charge formula in the vertex-transitive polyhedra with arbitrary genus, and calculate the corner charges in the simple model with genus $2$. 
In Sec.~\ref{Sec.5}, we show that the corner charge formulas in the vertex-transitive polyhedra are related to the two-dimensional disclination charges, and we establish a single universal formula for the disclination charge. 
In Sec.~\ref{sec.5.5}, by focusing on the disclination charge, we derive a formula for the coupling constant of the Wen-Zee action. 
Our conclusions are given in Sec.~\ref{Sec.6}. 

\section{Review of fractional corner charges in three-dimensional crystal shapes \label{Sec.2}}
\begin{figure}[t]
    \centering
    \includegraphics[scale=0.65]{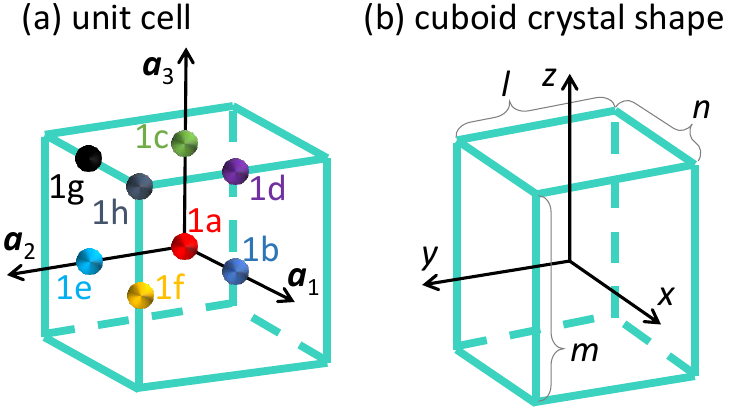}
    \caption{Calculation of the corner charge for the SG No.~47 ($Pmmm$) in the previous study \cite{PhysRevB.111.155305}. (a) Maximal WPs in the SG No.~47. $\bm{a}_{i}\ (i=1,2,3)$ are primitive lattice vectors in the orthorhombic unit cell. The cyan cuboid represents the unit cell. (b) Crystal shape of a cuboid with orthorhombic symmetry consisting of $l\times m\times n$ unit cells. $x$, $y$, and $z$ axes are parallel to $\bm{a}_{1}$, $\bm{a}_{2}$, and $\bm{a}_{3}$, respectively. }
    \label{fig:SGNo.47_unit_cell_and_crystal_shape}
\end{figure}

In this section, we briefly review the fractional corner charges in three-dimensional crystal shapes based on the previous works \cite{PhysRevB.109.085114,PhysRevB.111.155305}, where we construct a complete list of corner charge formulas for all the SGs and all the corresponding crystal shapes with a quantized corner charge. 
Here, for clarity of explanation, we consider the corner charge formula for the SG No.~47 ($Pmmm$), which corresponds to the orthorhombic point-group symmetry $D_{2h}$. 
The unit cell and the corresponding crystal shape are shown in Fig.~\ref{fig:SGNo.47_unit_cell_and_crystal_shape}(a) and (b), respectively. 

\subsection{Preliminary \label{Subsec.2-1}}
Before deriving the corner charge formulas for the SG No.~47, we discuss the setup required to describe the corner charges. 
First, we review the property of ionic and electronic charges in AIs. 
Electric charges in AIs consist of ions and electrons. 
The ionic charges in AIs are integer multiples of the elementary charge $\abs{e}$, where $-e~(e > 0)$ is the electron charge. 
Electrons in AIs occupy Wannier orbitals of electronic bands, which are exponentially localized, and the integral charges of electrons in the Wannier orbitals can be assigned to the Wannier centers. 
In the following, we assume that the gap is open both in the bulk and at the boundaries of AIs. 

Next, we also review the filling anomaly \cite{PhysRevB.99.245151,PhysRevB.103.165109}. 
In insulators of finite-sized crystals with some crystallographic symmetry, the charge neutrality for the whole crystal should be broken as long as the entire system is insulating, and the charge distribution fully reflects the crystallographic symmetry. 
In this case, the total charge of the entire system is called the filling anomaly. 
This filling anomaly is an integer and is a topological invariant.
We note that the nontrivial filling anomaly is seen in certain AIs called OAIs, which are characterized by charge imbalance between electrons and ions at each WP, and we hereafter focus on OAIs.  

Now, we discuss the filling anomaly in terms of maximal WPs. 
Here we focus on the crystal shape of a cuboid consisting of $l \times m \times n$ unit cells (Fig.~\ref{fig:SGNo.47_unit_cell_and_crystal_shape}(b)) for the SG No.~47, whose unit cell is shown in Fig.~\ref{fig:SGNo.47_unit_cell_and_crystal_shape}(a). 
Consider the number of occupied Wannier orbitals $n_{\omega}$ at a maximal WP $\omega~(= a,b,c,d,e,f,g,h)$. 
Let $m_{\omega}$ denote the total charge of ions measured in the unit of the elementary charge $\abs{e}$ at a maximal WP $\omega$. 
We then define the difference $\Delta \omega$ between them: 
\begin{align}
    \Delta \omega = n_{\omega} - m_{\omega}, 
\end{align}
where $\Delta \omega$ is always an integer by definition. 
We note that non-maximal WPs can be reduced to the maximal ones via continuous transformations, and thus we do not need to consider them. 

Thus, when calculating the filling anomaly, all we need to do is count the number of the values of $\Delta \omega$ at each WP in the unit cell shown in Fig.~\ref{fig:SGNo.47_unit_cell_and_crystal_shape}(a). 
To obtain the filling anomaly, we take the center of a crystal to be the maximal WP $1a$, and assume a perfect crystal, which implies that there is no surface reconstruction so that the periodicities of the crystal structure near the hinges and the surfaces reflect that of the bulk. 
Then the filling anomaly $\eta_{l,m,n}$ (per elementary charge) for the cuboid consisting of $l \times m \times n$ unit cells (Fig.~\ref{fig:SGNo.47_unit_cell_and_crystal_shape}(b)) can be expressed as a polynomial of $l$, $m$, and $n$, 
\begin{equation}
    \begin{aligned}[b]
        \eta_{l,m,n} =& \alpha_{7}lmn + \alpha_{6}lm + \alpha_{5}mn + \alpha_{4}ln \\
        &+ \alpha_{3}l + \alpha_{2}m + \alpha_{1}n + \Delta a, 
    \end{aligned}
\end{equation}
where $\alpha_{i}~(i = 1,\cdots,7)$ are integer constants.
Actually, $\alpha_{7}$ is proportional to the bulk charge density (per bulk unit cell), $\alpha_{i}~(i=4,5,6)$ are proportional to the surface charge densities (per surface unit cell) under the bulk charge neutrality condition $\alpha_{7}=0$.
Under the bulk and surface charge neutrality conditions $\alpha_{i}=0~(i=4,\cdots,7)$, $\alpha_{i}~(i=1,2,3)$ are proportional to the hinge charge densities (per hinge unit cell). 
Hereafter, we assume that the bulk, surfaces, and hinges are all charge neutral. 
In this case, the filling anomaly $\eta_{l,m,n}$ is given by $\Delta a$ in terms of mod $2$. 
Since all corners of the cuboid are equivalent under the point-group symmetry $D_{2h}$, which is the local symmetry at the WP $a$ in the SG No.~47, the corner charge formula $Q_{\text{corner}}$ for the SG No.~47 is given by the filling anomaly divided by the number of corners \cite{PhysRevB.111.155305},
\begin{align}
    Q_{\text{corner}} \equiv -\frac{\Delta a}{8}\abs{e}\ \ \left(\text{mod}\ \frac{\abs{e}}{4} \right), 
\end{align}
where we explicitly write the electron charge $-\abs{e}$.

\subsection{Real-space formulas of the corner charge in the vertex-transitive polyhedra with genus 0 \label{Subsec.2-2}}
\begin{figure}
    \centering
    \includegraphics[scale=0.52]{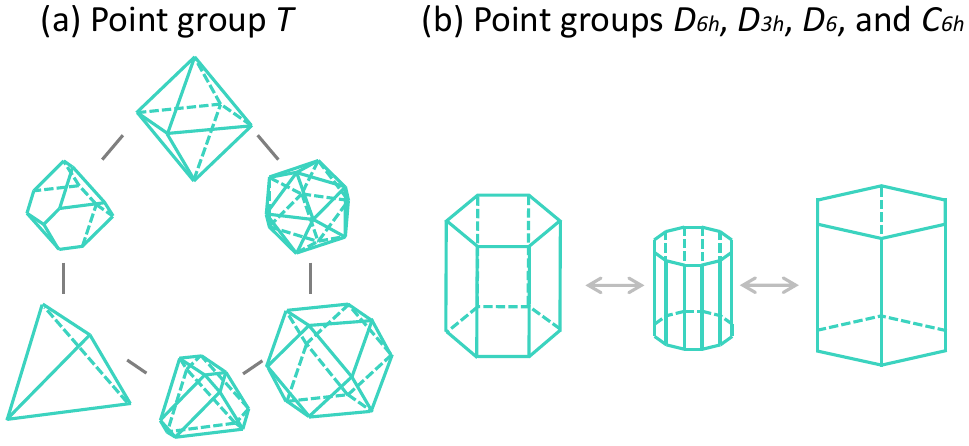}
    \caption{Vertex-transitive polyhedra corresponding to (a) the cubic point group $T$ in the spherical family, and (b) the hexagonal point groups $D_{6h}$, $D_{3h}$, $D_{6}$, and $C_{6h}$ in the cylindrical family. }
    \label{fig:previous_crystal_shapes}
\end{figure}

In order to obtain the corner charge formulas for all the SGs, and all the corresponding crystal shapes, we should consider the crystal shapes in more detail. 
For the quantization of the corner charge, we need to restrict the crystal shapes to be the vertex-transitive polyhedra \cite{polyhedra,bridges2002:320,ROBERTSON197079,https://doi.org/10.1112/jlms/s2-2.1.125,IsogonalPrismatoids}, in which all of the corners are equivalent by crystallographic symmetries, and thus the filling anomaly $Q_{\text{tot}}$ is divided equally among their corners. 
In the previous works \cite{PhysRevB.109.085114,PhysRevB.111.155305}, we restricted ourselves to the vertex-transitive polyhedra with genus $0$, and they are classified into the spherical family and the cylindrical family, whose examples are shown in Figs.~\ref{fig:previous_crystal_shapes}(a) and (b), respectively. 
For both families, the corner charge formulas $Q_{\text{corner}}$ in the three-dimensional crystals are given by the filling anomaly $Q_{\text{tot}}$ divided by the number of vertices $N$ as follows: 
\begin{align}
    Q_{\text{corner}} \equiv \frac{Q_{\text{tot}}}{N} = -\frac{\Delta a}{N}\abs{e}\ \ \left(\text{mod}\ \frac{2\abs{e}}{N}\right). \label{eq:global corner charge} 
\end{align}
Here we used the result of our previous works \cite{PhysRevB.109.085114,PhysRevB.111.155305} (Sec.~\ref{Sec.2}) that the filling anomaly $Q_{\text{tot}}$ is universally given by
\begin{align}
    Q_{\text{tot}} = -\Delta a \abs{e}\ \ (\text{mod}\ 2\abs{e}), \label{eq:delta_a_previous_work}
\end{align}
which is the total charge at the bulk WP $a$ located at the center of the crystal shapes.
Here, the WP $1a$ appears in the formula, because it is the WP located at the center of the crystal. 
We note that this result has the same form as in two dimensions \cite{PhysRevB.103.205123}, but unlike two dimensions, it is not known why the formula is same for all the SGs and for all the crystal shapes of vertex-transitive polyhedra. 

\section{Local form of the corner charge formulas for genus $0$ \label{Sec.3}} 
Here, on the right hand side of Eq.~(\ref{eq:global corner charge}), the number of corners $N$ is a global quantity, which depends on the entire shape of the crystal.
Meanwhile, the corner charge should physically be a local quantity. 
Therefore, we expect that the number of corners can be obtained from local information around the corners of the crystal shapes, which we derive in the following. 

\begin{figure}
    \centering
    \includegraphics[scale=0.6]{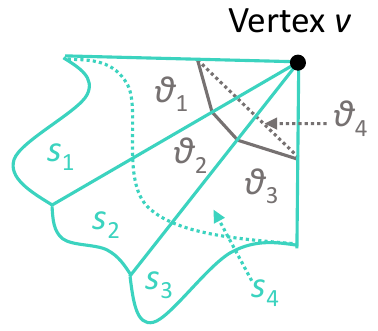}
    \caption{The corner $v$ in the polyhedron $A$. $s_{i}\ (i=1,2,3,4)$ is the $i$-th surface sharing the corner $v$. $\theta_{i}\ (i=1,2,3,4)$ is the interior angle at the $v$ on the surface $s_{i}$.}
    \label{fig:degree of sharpness}
\end{figure}
 
In three-dimensions, let $P$ be a three-dimensional vertex-transitive polyhedron, which represents the crystal shape, and let $A$ be its two-dimensional surface. 
In this section, we limit ourselves to the cases with a convex polyhedron with genus $0$. 
We focus on a certain corner $v$ in the polyhedron $A$. 
Then let $s_{i}\ (i=1,\cdots,n)$ be the $i$-th surface among the $n$ surfaces sharing the same corner $v$, and $\theta_{i}\ (i = 1,\cdots,n)$ be the interior angle at the corner $v$ on the surface $s_{i}$ as shown in Fig.~\ref{fig:degree of sharpness}. 
In this case, we can define the degree of sharpness $\delta_{v}$ of the corner $v$ as 
\begin{equation}
    \begin{aligned}[b]
        \delta_{v} &\equiv 2\pi - \sum_{i=1}^{n}\theta_{i} \\
        &=2\pi - \sigma_{v}, \label{eq:degree of sharpness}
    \end{aligned}
\end{equation}
where $\sigma_{v}=\sum_{i=1}^{n}\theta_{i}$. 
Since $A$ is a vertex-transitive polyhedron, the degrees of sharpness of all the vertices in $A$ have the same value: $\delta_{v} (\equiv \delta)$. 
Then, the number of corners $N$ in the polyhedron $A$ is given by 
\begin{align}
    N = \frac{4\pi}{\delta}. \label{eq:N}
\end{align}
From Eq.~(\ref{eq:N}), the corner charge in the three-dimensional crystal shapes is given as a local quantity by 
\begin{align}
    Q^{3D}_{\text{corner}} \equiv -\frac{\Delta a}{4\pi}\abs{e}\delta\ \ \left(\text{mod}\ \frac{\abs{e}}{2\pi}\delta\right). \label{eq:micro_corner_charge_3d}
\end{align}
From Eq.~(\ref{eq:micro_corner_charge_3d}), we find that the fractional corner charges can only be determined by the total charge $\Delta a$ at the center of the unit cell and the local geometry around the corners of the crystal shapes. 

In two dimensions, we also write the corner charge as a local quantity.
We note that the degree of sharpness $\delta$ in the two-dimensional crystal shapes with $C_{n}$ symmetry is given by $\delta\equiv\pi-\theta$, where $\theta$ is an interior angle, and is related to the number of corners $N$ by $N=\frac{2\pi}{\delta}$. 
Thus, the local expression of the corner charge formulas in the two-dimensional crystal shapes with $C_{n}$ symmetry is given by 
\begin{align}
    Q^{2D}_{\text{corner}} \equiv -\frac{\Delta a}{2\pi}\abs{e}\delta\ \ \left(\text{mod}\ \abs{e} \right). \label{eq:micro_corner_charge_2d}
\end{align}

\section{Corner charge formulas for arbitrary genus \label{Sec.4}}
In Sec.~\ref{Sec.3}, we derive the corner charge formulas for vertex-transitive polyhedra with genus $0$. 
On the other hand, since from Sec.~\ref{Sec.3}, the corner charge formulas are described in terms of local geometry around the corners, they may be applied to the cases with higher genus. 
In this section, we consider vertex-transitive polyhedra with arbitrary genus, and propose the corner charge formulas for them. 

\begin{figure}
    \centering
    \includegraphics[scale=0.55]{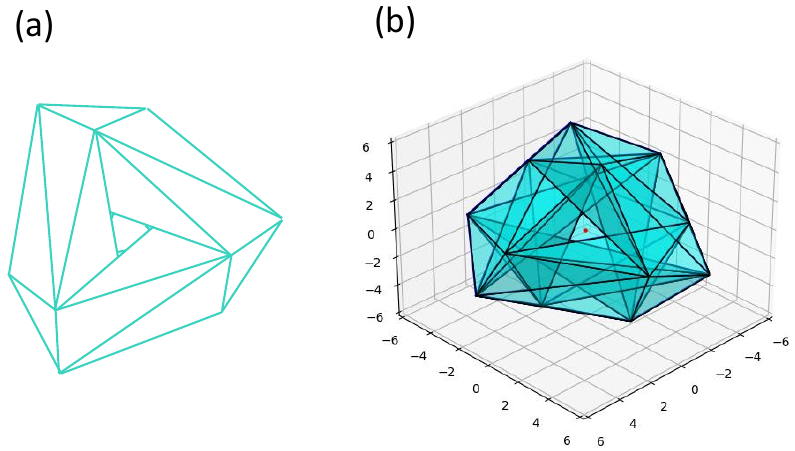}
    \caption{Vertex-transitive polyhedron with genus $3$ under the tetrahedral group of rotations \cite{Leopold}. }
    \label{fig:Vertex-transitive_polyhedra_genus3}
\end{figure}

\subsection{Corner charge formula for three-dimensional polyhedra with arbitrary genus \label{Subsec.4-1}}
The vertex-transitive polyhedra with higher genus are still largely unexplored mathematically, and the polyhedron shown in Fig.~\ref{fig:Vertex-transitive_polyhedra_genus3} is one of the examples introduced in Ref.~\cite{Leopold}. 
In such cases of higher genus, the following two points should be considered. 

First, in general cases with arbitrary genus, for a vertex-transitive polyhedron $A$, the number of vertices $N$ is expressed in terms of the degree of sharpness $\delta$, as $N=\frac{2\pi\chi(A)}{\delta}$, where $\chi(A)$ represents the Euler characteristic of $A$. 
The Euler characteristic $\chi(A)$ of $A$ is $\chi(A)=2-2g$ for genus $g$. 
Thus, we obtain
\begin{align}
    N = \frac{2\pi}{\delta}(2-2g), \label{eq:N_higher_genus}
\end{align} 
which reduces to Eq.~(\ref{eq:N}) for cases with genus $0$. 

Second, we revisit the definition of the filling anomaly. 
So far, we calculated the filling anomaly by counting the number of lattice points corresponding to the WP $a$ in the polyhedra. 
In this case, the Ehrhart polynominal \cite{beck2015computing,DiazRobins1996} is a useful tool to count the lattice points in the lattice polyhedra.

Let $P$ be a lattice polyhedron, in which all the vertices are located at the integer lattice points. 
Let $tP$ be the $t$-fold dilation of $P$, where $t$ is any positive integer. 
In this case, the number of lattice points in the polyhedron $tP$ including the boundaries is given by the Ehrhart polynominal $L(t,P)$: 
\begin{align}
    L(t,P) = \alpha_{3}t^{3} + \alpha_{2}t^{2} + \alpha_{1}t + \alpha_{0}, \label{eq:Ehrhart polynominal}
\end{align}
where the coefficients $\alpha_{i}\ (i=0,1,2,3)$ are rational numbers. 
Here the coefficient $\alpha_{0}$ is given by {\cite{beck2015computing, DiazRobins1996}
\begin{align}
    \alpha_{0} = \chi(P) = 1-g,
\end{align}
where $\chi(P)$ is the Euler characteristic. 
We note that since $P$ has the bulk region, the Euler characteristic $\chi(P)$ of $P$ is given by $1-g$, not by $2-2g\ (=\chi(A))$. 

Here we identify the WP $a$ with the lattice points. 
Then the coefficients $\alpha_{3},\alpha_{2},\alpha_{1}$ and $\alpha_{0}$ in Eq.~(\ref{eq:Ehrhart polynominal}) represent the charge densities for the bulk, surface, hinge, and corner regions.
Thus, under the charge neutrality conditions for the bulk, surfaces, and hinges (Sec.~\ref{Sec.2}A), $L(t,P)$ is equal to $\alpha_{0}\ (=1-g)$, and the filling anomaly $Q_{\text{total}}$ of $P$ with genus $g$ should be given by
\begin{align}
    Q_{\text{total}} = -\alpha_{0}\Delta a \abs{e} = -(1-g)\Delta a\abs{e}. \label{eq:general_filling_anomaly}
\end{align}

In particular, when the genus of $P$ is $0$, Eq.~(\ref{eq:general_filling_anomaly}) gives $\alpha_{0} = 1$ and $Q_{\text{total}} = -\Delta a\abs{e}$, as obtained in Eq.~(\ref{eq:delta_a_previous_work}) \cite{PhysRevB.109.085114,PhysRevB.111.155305}.
This result is natural, by noting that the Ehrhart polynominal $L(t,P)$ can be defined even for $t=0$. 
In this case, for a polyhedron with genus $0$, $tP$ consists only of the origin, and $L(0,P) = \alpha_{0}= 1$, which implies that corner charge formulas in three-dimensional crystal shapes with genus $0$ are universally given by the total charge at their center. 

To summarize, from Eqs.~(\ref{eq:N_higher_genus}) and (\ref{eq:general_filling_anomaly}), the corner charge $Q_{\text{corner}}$ for the three-dimensional vertex-transitive polyhedra $P$ with genus $g\ (g > 1)$ is given by 
\begin{align}
    Q_{\text{corner}} \equiv \frac{Q_{\text{total}}}{N} \equiv -\frac{\Delta a}{4\pi}\abs{e}\delta\ \ \left(\text{mod}\ \frac{\abs{e}}{2\pi}\abs{\delta} \right). \label{eq:corner charge formula higher genus}
\end{align} 
Remarkably, Eq.~(\ref{eq:corner charge formula higher genus}) is the same expression as the case of genus $0$ (Eq.~(\ref{eq:micro_corner_charge_3d})) regardless of its genus $g$. 
It is natural, because the corner charge should be a local quantity, and is determined only by $\delta$, which characterizes the shape of the corner.

\subsection{Corner charge formula for two-dimensional shell polyhedra}
For subsequent discussions, here we define vertex-transitive shell polyhedra.
Vertex-transitive shell polyhedra (VTSPs) are two-dimensional vertex-transitive structures consisting only of polygons in a three-dimensional space, as shown in Fig.~\ref{fig:vertex-transitive polyhedra with genus 2}. 
We note that the VTSPs include closed polyhedra such as regular polyhedra (e.g., Fig.~\ref{fig:vertex-transitive polyhedra with genus 2}(a)), and non-closed structures shown in Figs.~\ref{fig:vertex-transitive polyhedra with genus 2}(b), (c), (d), and (e), which can be obtained from the closed polyhedra by removing certain faces. 
In this case, for VTSPs, the validity of Eq.~(\ref{eq:Ehrhart polynominal}) to calculate the filling anomaly is unclear. 
Actually, we find that such a filling anomaly is universally given by 
\begin{align}
    Q_{\text{total}} = -(2-2g)\Delta a\abs{e}, \label{eq:filling anomaly without bulk}
\end{align}
whose proof is shown in Appendix.~\ref{Ap.proof}. 
In this case, $2-2g$ in Eq.~(\ref{eq:filling anomaly without bulk}) is equal to the Euler characteristic i.e., $\chi(R)=2-2g$ for the VTSP $R$ with genus $g$, and thus Eq.~(\ref{eq:Ehrhart polynominal}) is valid for the VTSPs.
By using Eq.~(\ref{eq:N_higher_genus}), the corner charge formulas of the VTSPs are universally given by 
\begin{align}
    Q_{\text{corner}} \equiv \frac{Q_{\text{total}}}{N} \equiv -\frac{\Delta a}{2\pi}\abs{e}\delta\ \ \left(\text{mod}\ \abs{e} \right). \label{eq:corner charge formula higher genus without bulk}
\end{align}

The ambiguity of the corner charge in terms of modulo $\abs{e}$ in Eq.~(\ref{eq:corner charge formula higher genus without bulk}) comes from "decoration" near the corners, i.e., deviation of the nuclei positions and the electronic states near the corners away from those in the bulk.
We note that there might also be decorations near the hinges, which can modify the modulo part, but here we assume them for a while.
This assumption is convenient in the comparison between the corner charge and the disclination charge in the next section. 
We will discuss the effect of the decorations near the hinges in Conclusion.

\subsection{Example: Fractional corner charge in the two-dimensional shell polyhedron with genus 2 \label{Subsec.4-2}}

\begin{figure}
    \centering
    \includegraphics[scale = 0.55]{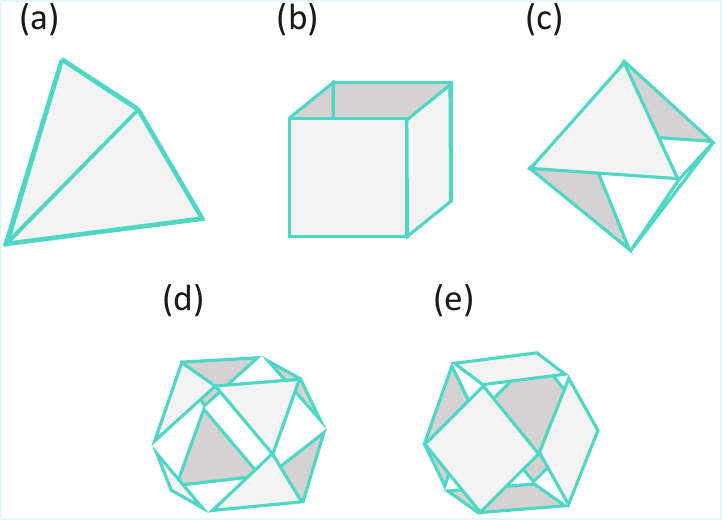}
    \caption{Examples of the VTSPs. (a) A regular tetrahedron. (b) A cubic structure with genus $1$. (c) A octahedral structure $R$ with genus $2$ (d) A cuboctahedral structure with genus $3$. (e) A cuboctahedral structure with genus $4$.}
    \label{fig:vertex-transitive polyhedra with genus 2}
\end{figure}

\begin{figure}[b]
    \centering
    \includegraphics[scale=0.6]{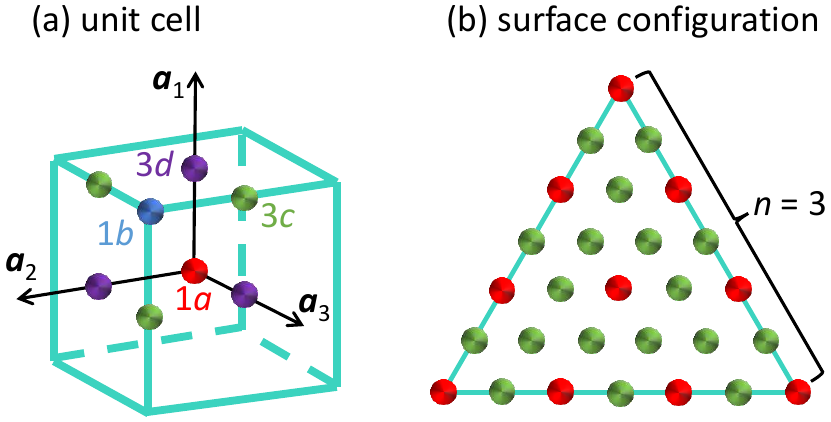}
    \caption{Calculation of the corner charge for the polyhedron $R$ in Fig.~\ref{fig:vertex-transitive polyhedra with genus 2} with the SG No.~195. (a) WPs in the SG No.~195. $\bm{a}_{i}\ (i=1,2,3)$ are primitive lattice vectors in the cubic unit cell. The cyan cube represents the unit cell. (b) Surface configuration of the polyhedron $R$ for the SG No.~195 when viewed from the [111] direction. WP $1a$ is located at each corner of the triangular surface of $R$ (Fig.~\ref{fig:vertex-transitive polyhedra with genus 2}(c)).}
    \label{fig:SG No.195_genus_2}
\end{figure}

In the previous subsections, we propose the corner charge formulas for the vertex-transitive polyhedra with higher genus as Eqs.~(\ref{eq:corner charge formula higher genus}) (3D polyhedra) and (\ref{eq:corner charge formula higher genus without bulk}) (2D shell polyhedra). 
In this section, we show an example to see the validity of our formulas. 
Consider the simple octahedral structure with genus $2$ as shown in Fig.~\ref{fig:vertex-transitive polyhedra with genus 2}(c). 
We find that this crystal shape, which we hereafter call $R$, is a VTSP with genus $2$ under the point-group symmetries $T$ or $T_{d}$.
Here, as an example, we assume that the bulk symmetry is the SG No.~195, whose point-group symmetry at the WP $1a$ is $T$. 
We assume that each corner of the polyhedron $R$ (Fig.~\ref{fig:vertex-transitive polyhedra with genus 2}(c)) is located at the WP $1a$. 
The filling anomaly of $R$ in Fig.~\ref{fig:SG No.195_genus_2}(b) (per elementary charge $\abs{e}$) is given by 
\begin{align}
    L(n,R) = 2n^{2}(\Delta a + 3\Delta c) + 6n(\Delta a + \Delta c) - 2\Delta a.
\end{align}
We find that the constant term agrees with its Euler characteristic $\chi(R) =2-2g= -2$ in Sec.~\ref{Sec.4}B. 
By considering the charge neutrality conditions for the bulk, and surfaces, i.e., $\Delta a + 3\Delta c = 0$ and $\Delta a + \Delta c \equiv 0\ \ (\text{mod}\ 1)$, respectively, the filling anomaly $L(n,R)$ (per elementary charge) is given by 
\begin{align}
    L(n,R) \equiv -2\Delta a \ \ (\text{mod}\ 6). 
\end{align}
In this case, since the sum $\sigma$ of the interior angles at a single corner of $R$ is $\sigma=\frac{2\pi}{3}$, the degree of sharpness $\delta$ is given by
\begin{equation}
    \begin{aligned}[b]
        \delta &\equiv -\sigma \\
        &= -\frac{2\pi}{3},
    \end{aligned}
\end{equation}
whose proof is shown in Appendix.~\ref{Ap.proof2}.
Thus, from Eq.~(\ref{eq:corner charge formula higher genus without bulk}), the corner charge formula $Q_{\text{corner}}$ of $R$ is given by 
\begin{align}
    Q_{\text{corner}} \equiv -\frac{\Delta a }{2\pi}\abs{e}\delta \equiv \frac{\Delta a}{3}\abs{e}\ \ (\text{mod}\ \abs{e}). \label{eq:corner_charge_genus 2}
\end{align}

This result is natural from the following consideration.
Because two vertices from two regular triangles (Fig.~\ref{fig:SG No.195_genus_2}(b)) merge into one corner in the polyhedron $R$ in Fig.~\ref{fig:vertex-transitive polyhedra with genus 2}(c), we expect that the result of Eq.~(\ref{eq:corner_charge_genus 2}) equals a double of the corner charge formula of the two-dimensional crystal shapes with $C_{3}$ symmetry.
It is indeed the case, since the corner charge of the two-dimensional crystal shapes with $C_{3}$ symmetry is given by $-\frac{\Delta a}{3}\abs{e}$ in terms of modulo $\abs{e}$ \cite{PhysRevB.103.205123}.  

Moreover, we note that there are 15 SGs, whose point-group symmetries at the WP $a$ are either $T$ or $T_{d}$, i.e., the SG numbers 195, 196, 197, 201, 203, 208, 210, 215, 216, 217, 218, 219, 224, 227, and 228. 
We derive the filling anomaly for those SGs as shown in Tab.~\ref{tab:filling anomaly genus 2}. 
We find that all the non-trivial filling anomalies are universally given by $-2\Delta a$, and those results support our proposal of the filling anomaly. 

We note that in SG numbers 196, 203, 210, 216, and 227, the filling anomalies are always trivial. 
In these SGs, the surface charge density and corner charge are proportional to the same charge imbalance $\Delta\omega$ at one certain WP, and therefore the corner charge formulas are always zero under the surface charge neutrality condition.

\begin{table}[ht]
    \centering
    \caption{Summary for point groups of WP $a$ and the filling anomalies for the 15 SGs. All the non-trivial filling anomalies are universally given by $-2\Delta a$.}
    \begin{ruledtabular}
        \begin{tabular}{ccc}
             SG number  & PG of WP $a$ & filling anomaly (mod 6) \\ \hline 
             195 & $T$ & $-2\Delta a$ \\
             196 & $T$ & $-2\Delta a(=0)$ \\
             197 & $T$ & $-2\Delta a$ \\
             201 & $T$ & $-2\Delta a$ \\
             203 & $T$ & $-2\Delta a(=0)$ \\
             208 & $T$ & $-2\Delta a$ \\
             210 & $T$ & $-2\Delta a(=0)$ \\ 
             215 & $T_{d}$ & $-2\Delta a$ \\
             216 & $T_{d}$ & $-2\Delta a(=0)$ \\
             217 & $T_{d}$ & $-2\Delta a$ \\
             218 & $T$ & $-2\Delta a$ \\
             219 & $T$ & $-2\Delta a$ \\
             224 & $T_{d}$ & $-2\Delta a$ \\
             227 & $T_{d}$ & $-2\Delta a(=0)$ \\
             228 & $T$ & $-2\Delta a$ \\
        \end{tabular}
    \end{ruledtabular}
    \label{tab:filling anomaly genus 2}
\end{table}

\section{Geometric relationship between disclination charge in two dimensions and corner charge in a two-dimensional polyhedron \label{Sec.5}}
Here we relate the corner charge of the crystal with the shape of VTSP in Sec.~\ref{Sec.4}B to the two-dimensional fractional disclination charge discussed in Refs.~\cite{PhysRevB.99.245151,PhysRevB.101.115115}.
We then show a general formulation of the disclination charges by using this relationship. 

\subsection{Review of fractional disclination charges in two dimensions \label{Subsec.5-1}}
Here we review the fractional disclination charges in two dimensions based on Ref.~\cite{PhysRevB.101.115115}. 
First, we briefly explain the concept of disclinations. 
A two-dimensional $C_{n}$ symmetric crystal can be divided into $n$ equivalent sectors, each of which has the internal angle $\frac{2\pi}{n}$ at the center of the crystal, and is mutually related by the $C_{n}$ symmetry. 
By adding or removing such sectors into/from the crystal, a zero-dimensional defect appears at the center of the crystal, which is called a disclination. 

The disclination is characterized by the holonomy for a closed path around its core, and the holonomy is denoted by the product of the translation and the rotation operators along the path. 
The rotation operator $\hat{r}(\Omega)$, which does not depend on the choice of the closed path, can be described by the Frank angle $\Omega$, while the translation operator $\hat{t}_{\bm{a}}$, which depends on the starting point of the closed path, can be described as $\bm{a} = a_{1}\bm{e}_{1} + a_{2}\bm{e}_{2}$, where $a_{i}\ (i = 1,2)$ are integers and $\bm{e}_{i}\ (i = 1,2)$ are primitive lattice vectors. 
Since the topological properties of the disclination are independent of the choice of the closed path, the classification of the disclination for a fixed Frank angle $\Omega = \pm\frac{2\pi}{n}$ is characterized by the conjugacy classes $[\bm{a}]^{(n)}$ of the holonomy group. 

\begin{figure}
    \centering
    \includegraphics[scale = 0.6]{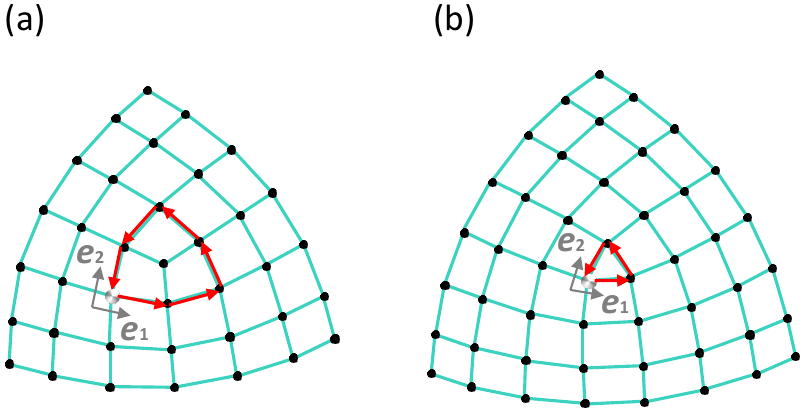}
    \caption{Two-dimensional $C_{4}$-symmetric square crystals with disclination of (a) $\Omega=-\frac{\pi}{2}$, $[\bm{a}]^{(4)}=0$ and (b) $\Omega=-\frac{\pi}{2}$, $[\bm{a}]^{(4)}=1$. The white dots represent the starting points of the loops. }
    \label{fig:disclination_e.g.}
\end{figure}

Consider the $C_{4}$-symmetric square lattice with disclination of $\Omega=-\frac{\pi}{2}$ as an example. 
In this case, there are two types of disclinations, i.e., $[\bm{a}]^{(4)} = 0$ in Fig.~\ref{fig:disclination_e.g.}(a) and $[\bm{a}]^{(4)}=1$ in Fig.~\ref{fig:disclination_e.g.}(b), in which the sum of $a_{1} + a_{2}$ from $\bm{a}=a_{1}\bm{e}_{1}+a_{2}\bm{e}_{2}$, taken over the vectors forming the closed path, is even or odd, respectively. 
Namely, the holonomy in Fig.~\ref{fig:disclination_e.g.}(a) is given by $\hat{r}(\frac{\pi}{2})\hat{t}_{2\bm{e}_{1}}\hat{r}(\frac{\pi}{2})\hat{t}_{2\bm{e}_{1}}\hat{r}(\frac{\pi}{2})\hat{t}_{2\bm{e}_{1}} = \hat{t}_{-2\bm{e}_{1}}\hat{r}(-\frac{\pi}{2})$, leading to $[\bm{a}]^{(4)}=0$, where we used $\hat{r}(R)\hat{t}_{\bm{a}}= \hat{t}_{R\bm{a}}\hat{r}(R)$, while the holonomy in Fig.~\ref{fig:disclination_e.g.}(b) is given by $\hat{r}(\frac{\pi}{2})\hat{t}_{\bm{e}_{1}}\hat{r}(\frac{\pi}{2})\hat{t}_{\bm{e}_{1}}\hat{r}(\frac{\pi}{2})\hat{t}_{\bm{e}_{1}} = \hat{t}_{-\bm{e}_{1}}\hat{r}(-\frac{\pi}{2})$, leading to $[\bm{a}]^{(4)}=1$. 
In the same way, $[\bm{a}]^{(2)}$ for the Frank angle $\Omega=\pi$ takes values in $\mathbb{Z}_{2}\oplus\mathbb{Z}_{2}$, corresponding to the cases with $a_{1}$ and $a_{2}$ being odd or even.
Likewise, $[\bm{a}]^{(3)}$ for the Frank angle $\Omega=\pm\frac{2\pi}{3}$ takes values in $\mathbb{Z}_{3}$, i.e., $a_{1}-a_{2}\ (\text{mod}\ 3)$, and $[\bm{a}]^{(6)}$ for the Frank angle $\Omega=\pm\frac{2\pi}{6}$ is zero. 

\begin{figure}
    \centering
    \includegraphics[scale = 0.6]{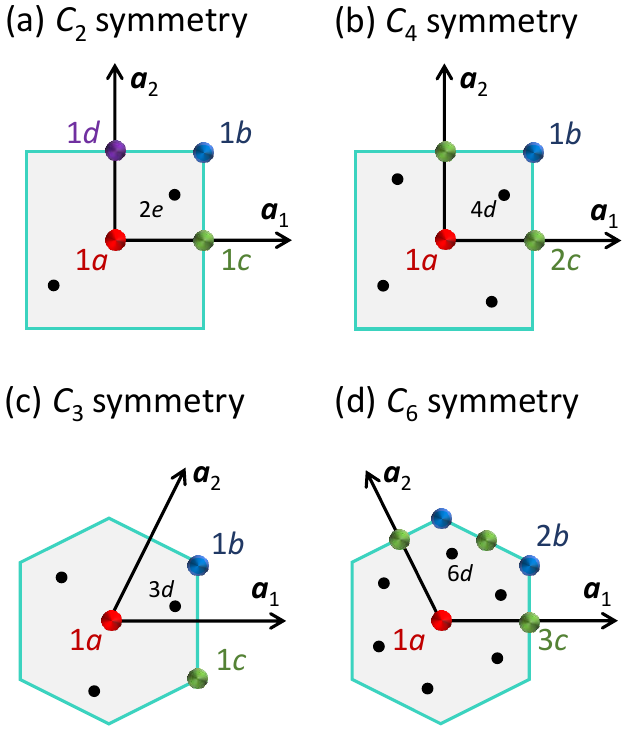}
    \caption{WPs in (a) $C_{2}$-, (b) $C_{4}$-, (c) $C_{3}$-, and (d) $C_{6}$-symmetric systems. The points with the same colors belong to the same WPs. $\bm{a}_{1}\ (=a\bm{e}_{1})$ and $\bm{a}_{2}\ (=a\bm{e}_{2})$ are primitive lattice vectors of lattice constant $a$. The gray regions are unit cells.  }
    \label{fig:2Dunitcell}
\end{figure}

\begin{figure*}
    \centering
    \includegraphics[scale=0.5]{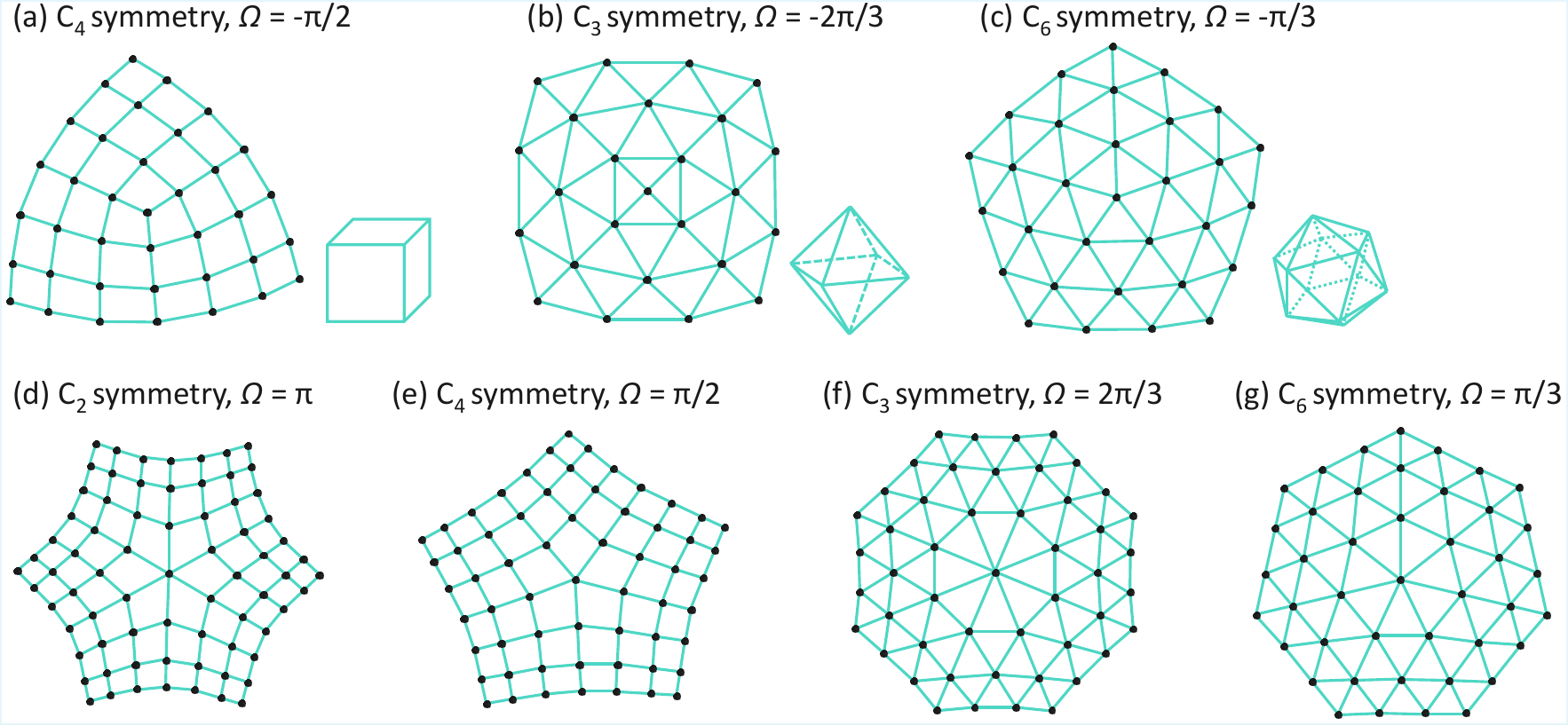}
    \caption{Two-dimensional disclination patterns with $[\bm{a}]^{(n)}=\bm{0}\ (n=2,4,3,6)$. (a) is $C_{4}$-symmetric lattice of Frank angle $\Omega=-\frac{\pi}{2}$. (b) is $C_{3}$-symmetric lattice of Frank angle $\Omega=-\frac{2\pi}{3}$. (c) is $C_{6}$-symmetric lattice of Frank angle $\Omega=-\frac{\pi}{3}$. (d) is $C_{2}$-symmetric lattice of Frank angle $\Omega=\pi$. (e) is $C_{4}$-symmetric lattice of Frank angle $\Omega=\frac{\pi}{2}$. (f) is $C_{3}$-symmetric lattice of Frank angle $\Omega=\frac{2\pi}{3}$. (g) is $C_{6}$-symmetric lattice of Frank angle $\Omega=\frac{\pi}{3}$. The polyhedra on the bottom right in (a), (b), and (c) show the corresponding polyhedra with the same degrees of sharpness as in the disclination cores in (a), (b), and (c), respectively.}
    \label{fig:disclination patterns}
\end{figure*}

Now we review the fractional disclination charges in two
dimensions \cite{PhysRevB.101.115115}. 
Consider two-dimensional $C_{n}$-symmetric systems, and let $\Delta \omega\ (=n_{\omega}-m_{\omega})$ denote the total charge of the ionic and electronic charges at the WP $\omega$ in the unit of $-|e|$ as discussed in Sec.~\ref{Sec.2}A. 
In this case, the bulk polarizations $\bm{P}^{(n)}\ (n=2,4,3,6)$ (per elementary charge $\abs{e}$) of the $C_{n}$-symmetric systems, whose unit cells are shown in Fig.~\ref{fig:2Dunitcell}, are given by \cite{PhysRevB.101.115115} 
\begin{equation}
    \begin{aligned}[b]
        \bm{P}^{(2)} &\equiv -\frac{\Delta b + \Delta c}{2}\bm{e}_{1} - \frac{\Delta b + \Delta d}{2}\bm{e}_{2}\ \ (\text{mod}\ 1), \\ 
        \bm{P}^{(4)} &\equiv -\frac{\Delta b + \Delta c}{2}(\bm{e}_{1} + \bm{e}_{2})\ \ (\text{mod}\ 1), \\
        \bm{P}^{(3)} &\equiv -\frac{\Delta b - \Delta c}{3}(\bm{e}_{1} + \bm{e}_{2})\ \ (\text{mod}\ 1), \\
        \bm{P}^{(6)} &\equiv 0. \label{eq:Ps}
    \end{aligned}
\end{equation} 
Then, the fractional disclination charges $Q^{(n)}_{\text{dis}}\ (n =2,4,3,6)$ in $C_{n}$-symmetric lattices with disclination of $\Omega$ (per elementary charge $\abs{e}$) are given by \cite{PhysRevB.101.115115}
\begin{equation}
    \begin{aligned}[b]
        Q^{(2)}_{\text{dis}} &\equiv -\frac{\Omega}{2\pi}(\Delta b + \Delta c + \Delta d) + \bm{T}^{(2)}\cdot\bm{P}^{(2)}\ \ (\text{mod}\ 1), \\
        Q^{(4)}_{\text{dis}} &\equiv -\frac{\Omega}{2\pi}(\Delta b + 2\Delta c)+ \bm{T}^{(4)}\cdot\bm{P}^{(4)}\ \ (\text{mod}\ 1), \\
        Q^{(3)}_{\text{dis}} &\equiv -\frac{\Omega}{2\pi}(\Delta b + \Delta c) + \bm{T}^{(3)}\cdot\bm{P}^{(3)}\ \ (\text{mod}\ 1),\\
        Q^{(6)}_{\text{dis}} &\equiv -\frac{\Omega}{2\pi}(2\Delta b + 3\Delta c)\ \ (\text{mod}\ 1). \label{eq:disclination charge formulas}
    \end{aligned}
\end{equation}
Here $\bm{T}^{(n)}\ \left(=a_{1}\bm{d}_{2}-a_{2}\bm{d}_{1}\right)$ is the vector perpendicular to the translation part of the holonomy, where $\bm{d}_{i}\cdot\bm{e}_{j}=\delta_{ij}$, and $\bm{T}^{(n)}\cdot\bm{P}^{(n)}\ (n=2,4,3)$ are given by
\begin{equation}
    \begin{aligned}[b]
        \bm{T}^{(2)}\cdot\bm{P}^{(2)} &\equiv -\frac{\Delta b + \Delta d}{2}a_{1} + \frac{\Delta b + \Delta c}{2}a_{2}\ \ (\text{mod}\ 1), \\
        \bm{T}^{(4)}\cdot\bm{P}^{(4)} &\equiv -\frac{\Delta b + \Delta c}{2}(a_{1}-a_{2}) \ \ (\text{mod}\ 1), \\
        \bm{T}^{(3)}\cdot\bm{P}^{(3)} &\equiv -\frac{\Delta b - \Delta c}{3}(a_{1} - a_{2})\ \ (\text{mod}\ 1). \label{eq:TP}
    \end{aligned}
\end{equation}
We note that the vector $\bm{T}^{(n)}$ depends on the choice of the closed path, while the vector $\bm{T}^{(n)}\cdot\bm{P}^{(n)}$ does not depend on it in terms of modulo $1$. 

\subsection{Relationship between fractional disclination charges and fractional corner charges in terms of geometric correspondence \label{Subsec.5-2}}
In Eqs.~(\ref{eq:disclination charge formulas}) and (\ref{eq:TP}), the disclination charge formulas have various forms.
Here, by identifying this disclination charge with a corner charge of a corresponding VTSP in Sec.~\ref{Sec.3}B, we show that the disclination charge formula in Eqs.~(\ref{eq:disclination charge formulas}) and (\ref{eq:TP}) is universally given by 
\begin{equation}
    Q^{(n)}_{\text{dis}} = \frac{\Omega}{2\pi}\Delta\omega\abs{e}\ \ (\text{mod}\ \abs{e}), \label{eq:general disclination formula}
\end{equation}
where $\omega$ is the WP at the disclination core. 
This WP $\omega$ depends on $[\bm{a}]^{(n)}$; $\omega=a$ when $[\bm{a}]^{(n)}=0$, and $\omega$ can be $b$, $c$, or $d$ when $[\bm{a}]^{(n)} \neq 0$.

To show Eq.~(\ref{eq:general disclination formula}), first, we restrict ourselves to the cases with $[\bm{a}]^{(n)}=0$ and $\Omega<0$ (Figs.~\ref{fig:disclination patterns}(a)--(c)).
We consider the $C_{4}$-symmetric lattice of Frank angle $\Omega=-\frac{\pi}{2}$ with $[\bm{a}]^{(4)}=0$ as shown in Fig.~\ref{fig:disclination patterns}(a). 
The three sectors with an angle $\theta=\frac{\pi}{2}$ share the disclination core $v_{\text{dis}}$, and from Eq.~(\ref{eq:degree of sharpness}), the degree of sharpness $\delta_{v_{\text{dis}}}$ is given by $2\pi-\frac{3\pi}{2}=\frac{\pi}{2}$, and $Q^{(4)}_{\text{dis}}$ is given by
\begin{align}
    Q^{(4)}_{\text{dis}} \equiv \frac{\Delta b + 2\Delta c}{4}\abs{e}\equiv -\frac{\Delta a}{4}\abs{e}\ \ \left(\text{mod}\ \abs{e}\right), \label{eq:disclination_charge_(b)}
\end{align}
from Eq.~(\ref{eq:disclination charge formulas}), where in the last equality, we used the charge neutrality condition for the bulk, i.e., $\Delta a + \Delta b + 2\Delta c =0$.
Since this value of $\delta$ is equal to that of a corner of the cube (without a bulk region), which is a VTSP, we expect that this disclination charge $Q_{\text{dis}}$ is equal to the corner charge of the cube, as we show below.
Here its unit cell is shown in Fig.~\ref{fig:2Dunitcell}(b). 
Since VTSP does not have the bulk region, the filling anomaly $Q_{\text{total}}$ is given by $-2\Delta a\abs{e}$ (Eq.~(\ref{eq:filling anomaly without bulk})). 
Thus, the fractional corner charge $Q_{\text{corner}}$ is given by 
\begin{equation}
    \begin{aligned}[b]
        Q_{\text{corner}} \equiv \frac{Q_{\text{total}}}{N} = -\frac{\Delta a}{4}\abs{e}\ \ \left(\text{mod}\ \abs{e}\right). \label{eq:corner_charge_(b)}
    \end{aligned}
\end{equation}
As expected, it is equal to the disclination charge $Q^{(4)}_{\text{dis}}$ in Eq.~(\ref{eq:disclination_charge_(b)}). 
Similarly, the disclination charge in Fig.~\ref{fig:disclination patterns}(b) ($\Omega=-\frac{2\pi}{3}$) is equal to the corner charge in the regular octahedron without the bulk region with its unit cell shown in Fig.~\ref{fig:2Dunitcell}(c). 
Likewise the disclination charge in Fig.~\ref{fig:disclination patterns}(c) ($\Omega=-\frac{\pi}{3}$) is equal to the corner charge in the regular icosahedron without the bulk region with its unit cell shown in Fig.~\ref{fig:2Dunitcell}(d).  
Thus in all the cases, the disclination charge is given by Eq.~(\ref{eq:general disclination formula}) with $\omega=a$, because the disclination core is at the WP $1a$. 

Now we consider other disclination patterns with $\Omega>0$, i.e., Figs.~\ref{fig:disclination patterns}(d), (e), (f) and (g). 
For example, in the $C_{2}$-symmetric lattice of the Frank angle $\Omega=\pi$ as shown in Fig.~\ref{fig:disclination patterns}(d), six sectors with an angle $\theta=\frac{\pi}{2}$ share the disclination core $v_{\text{dis}}$, and the degree of sharpness $\delta_{v_{\text{dis}}}$ is given by $2\pi-3\pi=-\pi$. 
In this case, from Eq.~(\ref{eq:disclination charge formulas}), the disclination charge $Q^{(2)}_{\text{dis}}$ with $[\bm{a}]^{(2)}=\bm{0}$ is given by
\begin{align}
    Q^{(2)}_{\text{dis}} \equiv -\frac{\Delta b + \Delta c +\Delta d}{2}\abs{e}\equiv \frac{\Delta a}{2}\abs{e}\ \ (\text{mod}\ \abs{e}), \label{eq:disclination_charge_(a)}
\end{align}
where in the last equality, we used the charge neutrality condition for the bulk, i.e., $\Delta a + \Delta b + \Delta c + \Delta d=0$.
Let us suppose that there exists a VTSP having the same value of $\delta (= -\pi)$ as above. 
Then, from Eq.~(\ref{eq:corner charge formula higher genus}), the fractional corner charge $Q_{\text{corner}}$ is given by
\begin{equation}
    \begin{aligned}[b]
        Q_{\text{corner}} &\equiv -\frac{\Delta a}{2\pi}\abs{e}\delta \equiv \frac{\Delta a}{2}\abs{e}\ \ (\text{mod}\ \abs{e}). \label{eq:corner_charge_(a)}
    \end{aligned}
\end{equation}
Eq.~(\ref{eq:corner_charge_(a)}) is equal to the disclination charge $Q^{(2)}_{\text{dis}}$ of $\Omega=\pi$ in Eq.~(\ref{eq:disclination_charge_(a)}), and Eq.~(\ref{eq:general disclination formula}) is satisfied. 
Likewise, we can show Eq.~(\ref{eq:general disclination formula}) also in the disclination shown in Figs.~\ref{fig:disclination patterns}(e), (f), and (g).

In the cases with $\Omega>0$, the corresponding polyhedra has genus $g > 0$.
Namely, the disclination charges in Figs.~\ref{fig:disclination patterns}(d), (e), (f) and (g) are equal to the corner charges in the VTSPs with genus $\frac{N}{4}+1$, $\frac{N}{8} + 1$, $\frac{N}{6} + 1$ and $\frac{N}{12} + 1$, respectively. 
Therefore, the disclination charges of the Frank angle $\Omega\ (>0)$ are equal to the corner charges in the corresponding VTSPs (with higher genus) having the same degree of sharpness as in the disclination core.
We note that existence of vertex-transitive polyhedra with higher genus consisting only of the single disclination pattern with the Frank angle $\Omega>0$ is a mathematically open question. 

So far we have considered the disclinations with $[\bm{a}]^{(n)}=\bm{0}$ shown in Fig.~\ref{fig:disclination patterns}. 
In such a case, as we have seen above, $\bm{T}^{(n)}\cdot\bm{P}^{(n)} \equiv \bm{0}\ (\text{mod}\ 1)\ (n=2,4,3)$ are satisfied, and the disclination charges are given by Eq.~(\ref{eq:general disclination formula}) with $\omega=a$, i.e., in terms of the total charge $\Delta a$ at the WP $1a$. 
On the other hand, even in the case of $[\bm{a}]^{(n)} \neq \bm{0}\ (n=2,4,3)$ (Figs.~\ref{fig:C4_disclination_patterns}, \ref{fig:C2_disclination_patterns}, and \ref{fig:C3_disclination_patterns}), Eq.~(\ref{eq:general disclination formula}) holds with $\omega=b$, $c$, or $d$.
In such cases, the center of the disclination patterns is occupied by some WPs $1\omega\ (\omega=b,c,d)$ other than WP $1a$ (see Appendix~\ref{Ap.calculation}), and the filling anomaly in the corresponding polyhedra can be described by the same $\Delta \omega$. 

Thus, to summarize, in all the possible disclination patterns (Figs.~\ref{fig:disclination patterns}, \ref{fig:C4_disclination_patterns}, \ref{fig:C2_disclination_patterns}, and \ref{fig:C3_disclination_patterns}), the disclination charge is given by Eq.~(\ref{eq:general disclination formula}), and is equal to the corner charge of the corresponding VTSP.
This formula (\ref{eq:general disclination formula}) is the unified formula of the disclination charge, which was written as Eqs.~(\ref{eq:disclination charge formulas}) and (\ref{eq:TP}) for the individual cases.

\section{Coupling constant in Wen-Zee action \label{sec.5.5}}
In the previous section, we show that the disclination charge is universally given by Eq.~(\ref{eq:general disclination formula}) by constructing the relationship between the two-dimensional disclination charges and the corner charges in the corresponding VTSPs. 
Meanwhile, two-dimensional disclination charges can be described by the Wen-Zee action \cite{PhysRevB.107.195153, PhysRevB.106.L241113}. 
Here we derive the value of the coupling constant in the Wen-Zee action by considering the disclination charge given by Eq.~(\ref{eq:general disclination formula}). 

First, we briefly review the Wen-Zee action. 
Let $M$ be a $(2 + 1)$-dimensional spacetime manifold: $M=\Gamma\times\mathbb{R}$, where $\Gamma$ is a two-dimensional spatial manifold and $\mathbb{R}$ is the time.
In this case, the Wen-Zee action represents the coupling between the gauge field $A\ (=A_{\mu}dx^{\mu})\ (\mu=0,1,2)$ and the background curvature $R$ in the manifold $\Gamma$ in the context of the quantum Hall effect \cite{PhysRevLett.69.953} as follows:
\begin{align}
    S_{\text{WZ}}[A,\omega,\bar{s}] = \frac{\bar{s}}{2\pi}\int_{M} A\wedge R = \frac{\bar{s}}{2\pi}\int_{M} A\wedge d\omega, 
\end{align}
where $\omega\ (=\omega_{\mu}dx^{\mu})\ (\mu=0,1,2)$ is the spin connection and $\bar{s}$ is the coupling constant. 
By considering the variation of the Wen-Zee action $S_{\text{WZ}}$ with respect to the scalar potential $A_{0}$, the excess charge density $\Delta\rho$ is obtained as
\begin{align}
    \Delta\rho = \frac{\delta S_{\text{WZ}}}{\delta A_{0}} = \frac{\bar{s}}{2\pi}d\omega, \label{eq:Wen-Zee rho}
\end{align}
where the factor $dx^{0}$ is omitted for simplicity.

Now, we consider a two-dimensional system with a disclination.
In this case, the curvature $R\ (=d\omega)$ is nonzero only at the disclination core, and given by $d\omega = \delta_{\text{dis}} =-\Omega$, where $\delta_{\text{dis}}$ is the degree of sharpness of the disclination core and $\Omega$ is the Frank angle. 
Thus, by integrating Eq.~(\ref{eq:Wen-Zee rho}) over space $\Gamma$ ($\partial\Gamma=0$), the fractional disclination charge $Q_{\text{dis}}$ is given by 
\begin{align}
    Q_{\text{dis}} =\int_{\Gamma}\Delta\rho =\frac{\bar{s}}{2\pi}\int_{\Gamma}d\omega = \frac{\bar{s}}{2\pi}\delta_{\text{dis}} = -\frac{\bar{s}}{2\pi}\Omega. \label{eq:field approach to discli}
\end{align}

By comparing Eq.~(\ref{eq:field approach to discli}) with Eq.~(\ref{eq:general disclination formula}), we conclude that the coupling constant $\bar{s}$ is given by
\begin{align}
    \bar{s} = -\Delta \omega\abs{e},
\end{align}
where $\omega$ is the WP at the disclination core.
When $[\bm{a}]^{(n)}=0$ as shown in Figs.~\ref{fig:disclination patterns}(a)--(g), $\omega$ is the WP $1a$, and when $[\bm{a}]^{(n)}\neq0$ (see Figs.~\ref{fig:C4_disclination_patterns}, \ref{fig:C2_disclination_patterns}, \ref{fig:C3_disclination_patterns}), $\omega$ is among the WPs $b$, $c$, and $d$.
Thus, we have derived the value of the coupling constant $\bar{s}$ in the Wen-Zee action. 

\section{Conclusion \label{Sec.6}}
In this paper, first we construct the corner charge formula as a local quantity by considering the geometrical property around the corners.
We find that the corner charge formulas are determined only by the filling anomaly $\Delta a$ and the degree of sharpness $\delta$, which is a local property of a corner, implying that our theory should also apply to the crystal shapes with higher genus. 
Here we assumed that the WP $1a$ is located at the center of the crystal.

Then, we extend the corner charge formulas also to the vertex-transitive polyhedra with arbitrary genus, by noting that the filling anomaly corresponds to the constant term of the Ehrhart polynominal.  
Surprisingly, the corner charge formulas for crystal shapes of three-dimensional polyhedra with arbitrary genera are universally given by Eq.~(\ref{eq:corner charge formula higher genus}), which is expressed in terms of $\delta$, confirming the local feature of the corner charge. 
Moreover, the corner charge formulas are given by Eq.~(\ref{eq:corner charge formula higher genus without bulk}) for the VTSPs, which are the structures consisting only of two-dimensional polygons.
We note that the classification of the vertex-transitive polyhedra with higher genus is a mathematically open question; meanwhile, our formulas apply also to those polyhedra with higher genus, which are yet to be found. 

As mentioned in Sec.~\ref{Sec.3}B, the decoration at the hinges may change the modulus of the corner charge in VTSPs in Eq.~(\ref{eq:corner charge formula higher genus without bulk}).
For example, in the tetrahedral shell polyhedron, modification of the electronic states at the hinges may change the corner charge by $\abs{e}/2$. 
This corresponds to attaching a non-trivial SSH chain with $\abs{e}/2$ charges at both ends onto every hinge.
Thus, it makes the modulo part in Eq.~(\ref{eq:corner charge formula higher genus without bulk}) to be $\abs{e}/2$.
Likewise, the corner charge will change by $h\times\frac{\abs{e}}{2}$ ($h$: the number of hinges meeting at each corners), by attaching a nontrivial SSH chain at each hinge.
Namely, in VTSPs with an odd number of $h$, the modulus of the corner charge in Eq.~(\ref{eq:corner charge formula higher genus without bulk}) will change from $\abs{e}$ to $\abs{e}/2$, as a result of possible decorations at hinges. 

Moreover, we show that the disclination charges in the two-dimensional systems are universally given by Eq.~(\ref{eq:general disclination formula}).
They equal to the corner charges in the corresponding VTSPs if they share the same degree of sharpness. 
In particular, we also find that for the disclination patterns with the Frank angle $\Omega\ (>0)$, the corresponding polyhedra has higher genus ($g>0$). 
We note that the corresponding VTSPs are determined only by the degree of sharpness for the disclination center.
In that case, all the vertices in the VTSP should be connected by certain three-dimensional point group symmetry due to the vertex transitiveness of the VTSP.

Finally, we discuss the relationship between our disclination charge formula (Eq.~(\ref{eq:general disclination formula})) and  the Wen-Zee action.
In this case, we show that the coupling constant $\bar{s}$ of the Wen-Zee action is given by the total charge $-\Delta \omega\abs{e}$ at the WP $\omega$ where $\omega$ is the WP at the center of the disclination. 

\begin{figure}[b]
    \centering
    \includegraphics[scale=0.5]{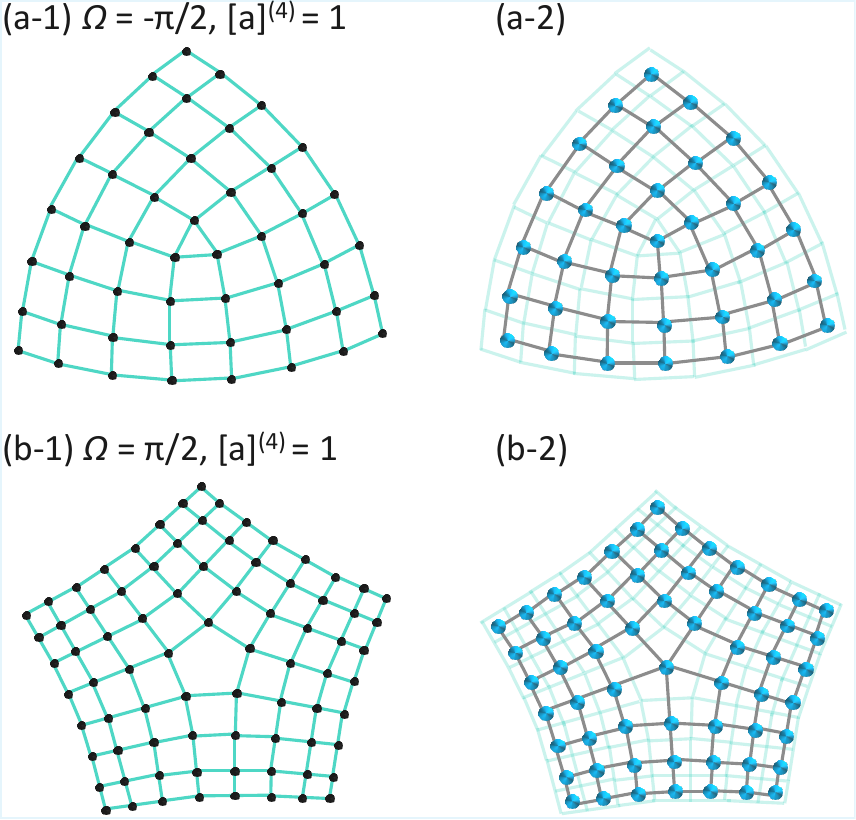}
    \caption{Two-dimensional $C_{4}$-symmetric lattices with disclination of (a-1) $\Omega=-\frac{\pi}{2}$, $[\bm{a}]^{(4)}=1$ and (b-1) $\Omega=\frac{\pi}{2}$, $[\bm{a}]^{(4)}=1$. (a-2) is the configuration of WP $b$ in (a-1). (b-2) is the configuration of WP $b$ in (b-1). Blue dots are WP $b$.}
    \label{fig:C4_disclination_patterns}
\end{figure}

\begin{figure*}[ht]
    \centering
    \includegraphics[scale=0.5]{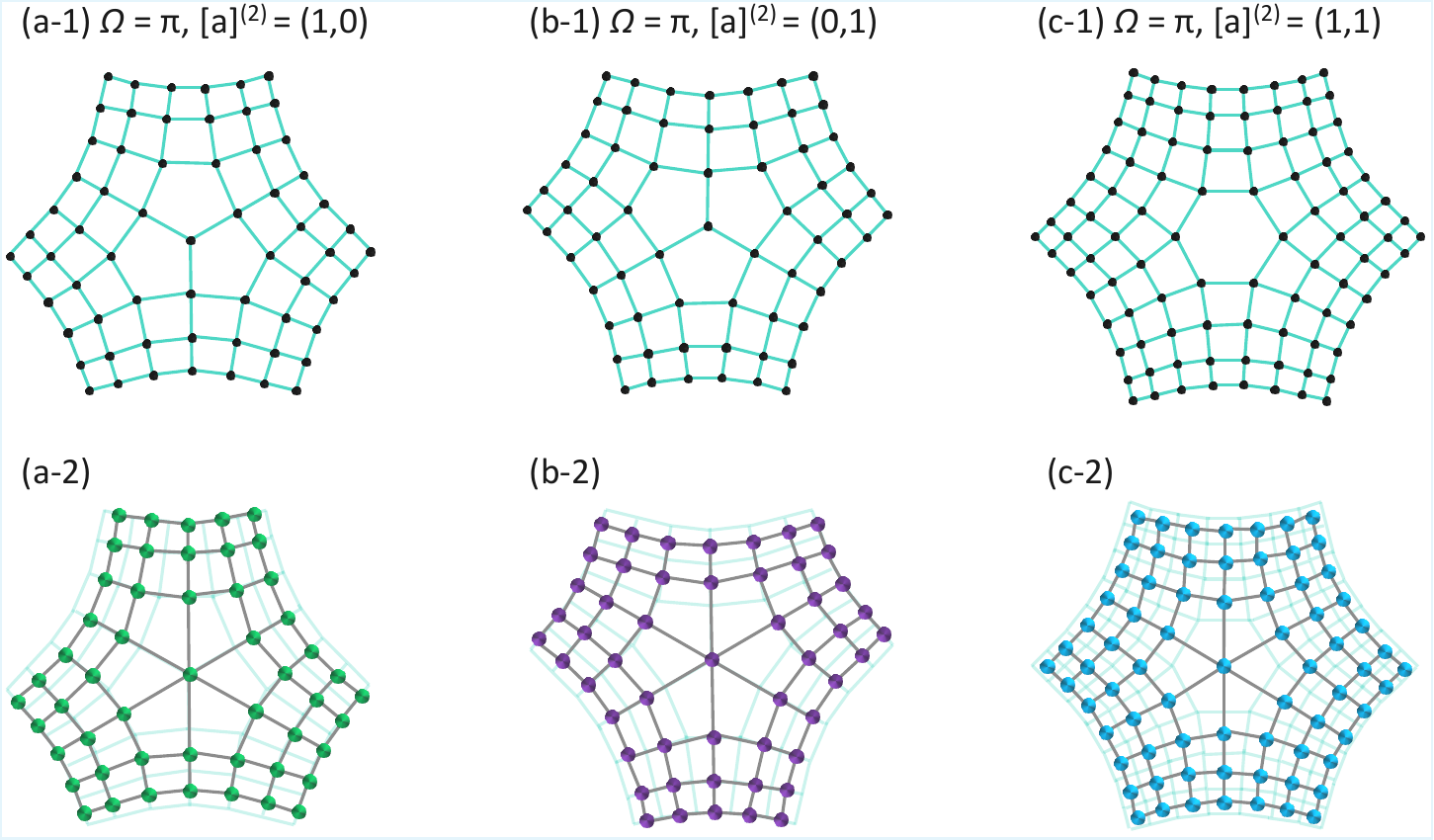}
    \caption{Two-dimensional $C_{2}$-symmetric lattices with disclination of (a-1) $\Omega=\pi$, $[\bm{a}]^{(2)}=(1,0)$, (b-1) $\Omega=\pi$, $[\bm{a}]^{(2)}=(0,1)$, and (c-1) $\Omega=\pi$, $[\bm{a}]^{(2)}=(1,1)$. (a-2) is the configuration of WP $c$ in (a-1). (b-2) is the configuration of WP $d$ in (b-1). (c-2) is the configuration of WP $b$ in (c-1). Green, purple and blue dots are the WP $c$, $d$ and $b$, respectively. }
    \label{fig:C2_disclination_patterns}
\end{figure*}

\begin{acknowledgments}

We thank S. Ono and K. Shiozaki for useful comments.
H. W. is supported by Japan Society for the Promotion of Science (JSPS) KAKENHI Grant No.~25KJI274.
S. M. is supported by Japan Society for the Promotion of Science (JSPS) KAKENHI Grant No.~JP22H00108, JP22K18687, and JP24H02231. 
\end{acknowledgments}

\appendix
\section{Filling anomaly of the shell polyhedra \label{Ap.proof}}
Here, we show that the filling anomaly in the two-dimensional shell polyhedra is universally given by $Q_{\text{total}} = -(2-2g)\Delta a\abs{e}$ (Eq.~(\ref{eq:filling anomaly without bulk})). 
Let $P$ be a polyhedron with genus $g$ consisting of $F$ surfaces, $E$ edges, and $V$ vertices. 
In this case, the polyhedron $P$ satisfies $F + V - E = 2(1-g)$ by Euler's theorem. 
Moreover, we define $L(n,\partial P)$ and $L(n,\partial^{2} P)$ as the number of lattice points on the surface $\partial P$ of $P$ including the boundaries and the number of lattice points on the edge $\partial^{2} P$ of $P$ including the end points, respectively. 
In this case, from the Ehrhart polynominal, $L(n,\partial P)$ and $L(n,\partial^{2} P)$ are given by 
\begin{align}
    L(n,\partial P) &= \alpha_{2}n^{2} + \alpha_{1}n + 1, \label{eq:appendix_1} \\
    L(n,\partial^{2} P) &= \alpha^{\prime}_{1}n + 1, \label{eq:appendix_2}
\end{align}
respectively, where $n$ is the length of an edge of $P$ (per unit cell), and $\alpha_{1}$, $\alpha_{2}$ and $\alpha^{\prime}_{1}$ are constants. 
The number of lattice points $L(n,P)$ in $P$ including the boundaries is the sum of the surface parts, the edge parts and the corner parts as follows:
\begin{equation}
    \begin{aligned}[b]
        L(n,P) =& \left(L(n,\partial P) - \frac{2E}{F}(L(n,\partial^{2} P)-1)\right)F \\
        & + (L(n,\partial^{2} P) - 2)E + V \\
        =& L(n,\partial P)F - L(n,\partial^{2} P)E + V \\
        =& \alpha_{2}Fn^{2} + (\alpha_{1}F - \alpha^{\prime}_{1}E)n + F + V - E \\
        =& \alpha_{2}Fn^{2} + (\alpha_{1}F - \alpha^{\prime}_{1}E)n + (2-2g), \label{eq:appendix_3}
    \end{aligned}
\end{equation}
where we used Eqs.~(\ref{eq:appendix_1}) and (\ref{eq:appendix_2}). 
Thus, from Eq.~(\ref{eq:appendix_3}), the filling anomaly of $P$ is given by 
\begin{align}
    Q_{\text{total}} = -(2-2g)\Delta a\abs{e},
\end{align}
where we assume that the WP $a$ is located at the center of the unit cell of $P$. 

\section{Degree of sharpness for the non-closed vertex-transitive shell polyhedra \label{Ap.proof2}}
There are various types of VTSPs, as shown in Fig.~\ref{fig:vertex-transitive polyhedra with genus 2}.
For closed VTSPs as shown in Fig~\ref{fig:vertex-transitive polyhedra with genus 2}(a), the degree of sharpness $\delta$ is given by Eq.~(\ref{eq:degree of sharpness}).
Here, we focus on the non-closed VTSPs, as shown in Figs.~\ref{fig:vertex-transitive polyhedra with genus 2}(c), (d) and (e), and show that the degree of sharpness $\delta^{\prime}$ for those VTSPs is $\delta^{\prime} = -\sigma$, where $\sigma$ is the sum of the interior angles at a single corner of the VTSPs. 

First, we show that the number of edges incident to a single corner in the closed vertex-transitive polyhedra with genus $0$ is strictly less than six. 
Since each surface is formed by at least three edges and each edge is shared by two surfaces, the number of surfaces $F$ and the number of edges $E$ in the VTSP with genus $0$ satisfy the following inequality:
\begin{align}
    3F \leq 2E. \label{eq:appendix_ineq1}
\end{align}
Then by using the Euler's theorem $F + V - E = 2$, where $V$ is the number of vertices, Eq.~(\ref{eq:appendix_ineq1}) is written by 
\begin{align}
    6 \leq 3V-E. \label{eq:appendix_ineq2}
\end{align}
Let $e$ be the number of edges incident to a single corner of the polyhedra.
Then the number of edges $E$ is given by $E=\frac{eV}{2}$. 
In this case, Eq.~(\ref{eq:appendix_ineq2}) is written as 
\begin{align}
    12 \leq (6-e)V.
\end{align}
Thus, it is shown that the number of edges $e$ incident to a single corner satisfies the following inequality:
\begin{align}
    (3\leq)\ e < 6. \label{eq:appendix_ineq3}
\end{align}

Now we derive the degree of sharpness $\delta^{\prime}$ of the VTSPs as shown in Figs.~\ref{fig:vertex-transitive polyhedra with genus 2}(c), (d), and (e).
Here, we assume that the non-closed VTSPs are generated by removing some faces from the closed VTSPs with the genus $0$ so that the number of edges $e$ incident to a single corner in the VTSPs with the higher genus satisfies Eq.~(\ref{eq:appendix_ineq3}).
Moreover, we assume that the surfaces constituting the non-closed VTSPs intersect only at the vertices. 
In this case, since the number of edges $e$ incident to a single corner of the VTSPs should be even, from Eq.~(\ref{eq:appendix_ineq3}), $e$ is $4$, which means that there are two surfaces sharing a single corner of the VTSPs. 

Let $R^{\prime}$ be the VTSP with genus $g$ satisfying the above assumptions, consisting of $F^{\prime}$ surfaces, $E^{\prime}$ edges, and $V^{\prime}$ vertices. 
In this case, the VTSP $R^{\prime}$ consists of $F^{\prime}_{1}$ regular $n_{1}$-gons and $F^{\prime}_{2}$ regular $n_{2}$-gons, and satisfies the following equations:
\begin{align}
    F^{\prime}&=F^{\prime}_{1}+F^{\prime}_{2}, \label{eq:appendix_sumF} \\
    E^{\prime}=2V^{\prime}&=n_{1}F^{\prime}_{1}+n_{2}F^{\prime}_{2}.
    \label{eq:appendix_EandV}
\end{align}
Then by considering the sum of all the interior angles for $R^{\prime}$, the sum $\sigma$ of the interior angles at a single corner of the VTSP is given by 
\begin{equation}
    \begin{aligned}[b]
        \sigma &= \frac{\pi}{V^{\prime}}\left((n_{1}-2)F_{1}^{\prime} + (n_{2}-2)F_{2}^{\prime}\right) \\
        &= \frac{2\pi}{V^{\prime}}\left(V^{\prime}-F^{\prime}\right), \label{eq:appendix_sigma}
    \end{aligned}
\end{equation}
where in the last equality, we use Eqs.~(\ref{eq:appendix_sumF}) and (\ref{eq:appendix_EandV}).
On the other hand, the degree of sharpness $\delta^{\prime}$ for $R^{\prime}$ is given by
\begin{equation}
    \begin{aligned}[b]
        \delta^{\prime} &= \frac{2\pi(2-2g)}{V^{\prime}} \\
        &= \frac{2\pi}{V^{\prime}}\left(F^{\prime}-V^{\prime}\right),\label{eq:appendix_delta}
    \end{aligned}
\end{equation}
where in the last equality, we use the Euler's theorem $F^{\prime} + V^{\prime} - E^{\prime} = 2-2g$ and Eq.~(\ref{eq:appendix_EandV}).
Thus, from Eqs.~(\ref{eq:appendix_sigma}) and (\ref{eq:appendix_delta}), the degree of sharpness $\delta^{\prime}$ is given by
\begin{align}
    \delta^{\prime} = -\sigma.
\end{align}

\begin{figure*}[t]
    \centering
    \includegraphics[scale=0.48]{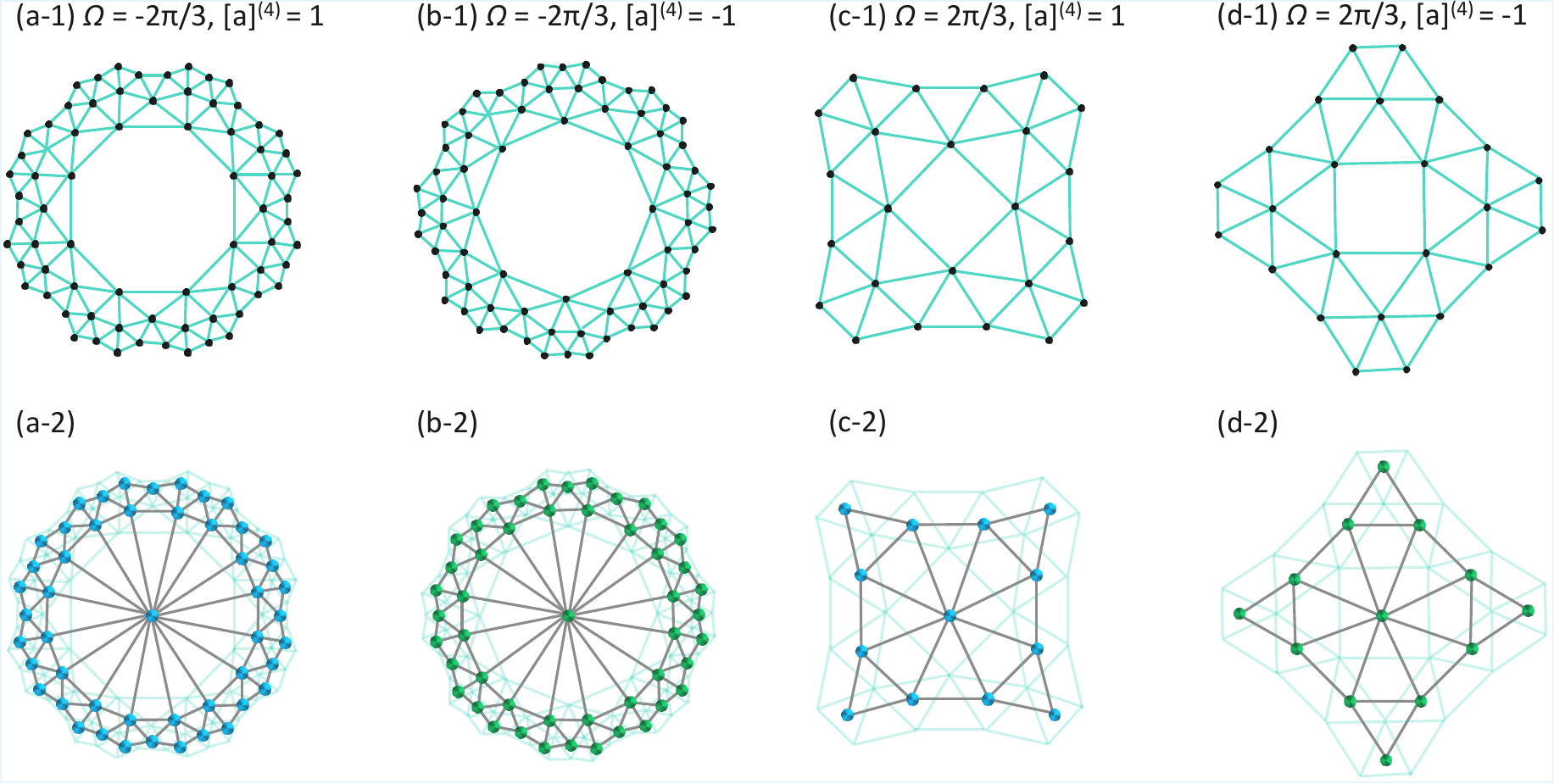}
    \caption{Two-dimensional $C_{3}$-symmetric lattices with disclination of (a-1) $\Omega=-\frac{2\pi}{3}$, $[\bm{a}]^{(3)}=1$, (b-1) $\Omega=-\frac{2\pi}{3}$, $[\bm{a}]^{(3)}=-1$, (c-1) $\Omega=\frac{2\pi}{3}$, $[\bm{a}]^{(3)}=1$, and (d-1) $\Omega=\frac{2\pi}{3}$, $[\bm{a}]^{(2)}=-1$. (a-2) and (c-2) are the configurations of WP $b$ in (a-1) and (c-1), respectively. (b-2) and (d-2) are the configurations of WP $c$ in (b-1) and (d-1), respectively. Blue and green dots are the WP $b$ and $c$, respectively.}
    \label{fig:C3_disclination_patterns}
\end{figure*}

\section{General correspondence between the two-dimensional disclination charges and three-dimensional corner charges \label{Ap.calculation}}
In Subsec.~\ref{Sec.6}B, we established the universal formula Eq.~(\ref{eq:general disclination formula}) for the disclination charge, and constructed the correspondence between the two-dimensional disclination charges and the three-dimensional corner charges for the disclination patterns with $[\bm{a}]^{(n)} = \bm{0}\ (n=2,4,3)$, where $\omega$ is given by $\omega=a$.
Here, we show that this correspondence is generally satisfied even for the cases of $[\bm{a}]^{(n)} \neq \bm{0}\ (n=2,4,3)$, where $\omega$ is given either by $b$, $c$, or $d$ as shown in Figs.~\ref{fig:C4_disclination_patterns}, \ref{fig:C2_disclination_patterns}, and \ref{fig:C3_disclination_patterns}. 

\subsection{$C_{4}$-symmetric lattice}
We consider the $C_{4}$-symmetric lattice with the Frank angle $\Omega=-\frac{\pi}{2}$ and $[\bm{a}]^{(4)}=1$ as shown in Fig.~\ref{fig:C4_disclination_patterns}(a-1), where the lattice points representing the WP $a$ (Fig.~\ref{fig:2Dunitcell}(b)) are plotted. 
By plotting the WP $b$ as the lattice point instead of the WP $a$, we get Fig.~\ref{fig:C4_disclination_patterns}(a-2), which is the same as Fig.~\ref{fig:disclination patterns}(a), showing the WP $a$ of the $C_{4}$-symmetric lattice with $\Omega = -\delta = -\frac{\pi}{2}$, and $[\bm{a}]^{(4)}=0$. 
Thus, from Eq.~(\ref{eq:corner_charge_(b)}), the fractional corner charge $Q_{\text{corner}}$ in the corresponding polyhedron (i.e., a cube) is given by 
\begin{equation}
    \begin{aligned}[b]
        Q_{\text{corner}} \equiv \frac{Q_{\text{total}}}{N} = -\frac{\Delta b}{4}\abs{e}\ \ \left(\text{mod}\ \abs{e}\right), \label{eq:APP B C4 -pi/2} 
    \end{aligned}
\end{equation}
which is the same as Eq.~(\ref{eq:general disclination formula}) with $\omega=b$. 
Indeed, Eq.~(\ref{eq:disclination charge formulas}) with $\Omega=-\frac{\pi}{2}$ and $[\bm{a}]^{(4)}=1$ gives the disclination charge $Q^{(4)}_{\text{dis}}$ equal to Eq.~(\ref{eq:APP B C4 -pi/2}).

In the same way, the $C_{4}$-symmetric lattice with $\Omega=\frac{\pi}{2}$ and $[\bm{a}]^{(4)}=1$ is shown in Fig.~\ref{fig:C4_disclination_patterns}(b-1).
When the lattice points at the WP $b$ are plotted, the resulting disclination pattern (Fig.~\ref{fig:C4_disclination_patterns}(b-2)) is the same as Fig.~\ref{fig:disclination patterns}(e) for the $C_{4}$-symmetric lattice with $\Omega = -\delta = \frac{\pi}{2}$, and $[\bm{a}]^{(4)}=0$. 
Thus, the corresponding polyhedron is the VTSP with genus $\frac{N}{8}+1$, and the fractional corner charge $Q_{\text{corner}}$ is given by 
\begin{equation}
    \begin{aligned}[b]
        Q_{\text{corner}} \equiv \frac{Q_{\text{total}}}{N} = \frac{\Delta b}{4}\abs{e}\ \ \left(\text{mod}\ \abs{e}\right),
    \end{aligned}
\end{equation}
which is equal to the disclination charge in Eq.~(\ref{eq:general disclination formula}) with $\omega=b$.

\subsection{$C_{2}$-symmetric lattice}
We consider the $C_{2}$-symmetric lattices with the Frank angle $\Omega=\pi$ as shown in Figs.~\ref{fig:C2_disclination_patterns}(a-1), (b-1), and (c-1) ($[\bm{a}]^{(2)}=(1,0)$, $[\bm{a}]^{(2)}=(0,1)$, and $[\bm{a}]^{(2)}=(1,1)$, respectively). 
For (a-1), (b-1), and (c-1) in Figs.~\ref{fig:C2_disclination_patterns}, we reconstruct the disclination patterns as shown in Figs.~\ref{fig:C2_disclination_patterns}(a-2), (b-2), and (c-2), respectively, by adopting the WP $c$ for (a-1), $d$ for (b-1), and $b$ for (c-1) as the lattice points.
Except for the difference in the WPs corresponding to the lattice points, we find that all the disclination patterns (Figs.~\ref{fig:C2_disclination_patterns}(a-2), (b-2), and (c-2)) are the same as Fig.~\ref{fig:disclination patterns}(d) showing the $C_{2}$-symmetric lattice with $\Omega=-\delta=\pi$ and $[\bm{a}]^{(2)}=\bm{0}$. 
Thus, from Eq.~(\ref{eq:corner charge formula higher genus without bulk}), the fractional corner charges $Q_{\text{total}}$ for the cases of Figs.~\ref{fig:C2_disclination_patterns}(a-2), (b-2), and (c-2) are given by
\begin{align}
    Q^{\text{(a-2)}}_{\text{corner}} = \frac{\Delta c}{2}\abs{e}\ \ \left(\text{mod}\ \abs{e}\right), \\
    Q^{\text{(b-2)}}_{\text{corner}} = \frac{\Delta d}{2}\abs{e}\ \ \left(\text{mod}\ \abs{e}\right), \\
    Q^{\text{(c-2)}}_{\text{corner}} = \frac{\Delta b}{2}\abs{e}\ \ \left(\text{mod}\ \abs{e}\right), 
\end{align}
respectively. 
They are equal to the disclination charges in Eq.~(\ref{eq:general disclination formula}) with $\omega=c,\ d,\ b$, respectively.

\subsection{$C_{3}$-symmetric lattice}
Finally, we consider the $C_{3}$-symmetric lattices with the Frank angles $\Omega=-\frac{2\pi}{3}$ (Figs.~\ref{fig:C3_disclination_patterns}(a-1) and (b-1)) and $\frac{2\pi}{3}$ (Figs.~\ref{fig:C3_disclination_patterns}(c-1) and (d-1)). 
In the same way as in the $C_{4}$- and $C_{2}$-symmetric cases, reconstructions of the disclination patterns for (a-1), (b-1), (c-1), and (d-1) in Figs.~\ref{fig:C3_disclination_patterns} are shown in Figs.~\ref{fig:C3_disclination_patterns}(a-2), (b-2), (c-2), and (d-2), respectively, which are all $[\bm{a}]^{(3)}=0$. 

For the configurations shown in Figs.~\ref{fig:C3_disclination_patterns}(a-2) and (b-2), we designate the WPs $b$ and $c$ as the lattice points, respectively. 
In these cases, 16 angles with $\theta=\frac{\pi}{3}$ share the disclination cores, and the degree of sharpness $\delta$ of the corresponding polyhedra is given by $\delta=2\pi-\frac{16\pi}{3}=-\frac{10\pi}{3}$. 
Thus, from Eq.~(\ref{eq:corner charge formula higher genus without bulk}), the fractional corner charges $Q_{\text{corner}}$ for the cases of Figs.~\ref{fig:C3_disclination_patterns}(a-2) and (b-2) are given by
\begin{align}
    Q^{\text{(a-2)}}_{\text{corner}} =  \frac{5\Delta b}{3}\abs{e} \equiv -\frac{\Delta b}{3}\abs{e}\ \ (\text{mod}\ \abs{e}), \\
    Q^{\text{(b-2)}}_{\text{corner}} = \frac{5\Delta c}{3}\abs{e} \equiv -\frac{\Delta c}{3}\abs{e}\ \ (\text{mod}\ \abs{e}) ,
\end{align}
respectively. 
They are equal to the disclination charges with $\omega=b,\ c$ in Eq.~(\ref{eq:general disclination formula}).

For the configurations shown in Figs.~\ref{fig:C3_disclination_patterns}(c-2) and (d-2), we designate the WPs $b$ and $c$ as the lattice points, respectively. 
In these cases, eight angles with $\theta=\frac{\pi}{3}$ share the disclination cores, and the degree of sharpness $\delta$ of the corresponding polyhedra is given by $\delta=2\pi-\frac{8\pi}{3}=-\frac{2\pi}{3}$. 
Thus, from Eq.~(\ref{eq:corner charge formula higher genus without bulk}), the fractional corner charges $Q_{\text{corner}}$ for the cases of Figs.~\ref{fig:C3_disclination_patterns}(c-2) and (d-2) are given by
\begin{align}
    Q^{\text{(c-2)}}_{\text{corner}} = \frac{\Delta b}{3}\abs{e}\ \ (\text{mod}\ \abs{e}), \\
    Q^{\text{(d-2)}}_{\text{corner}} = \frac{\Delta c}{3}\abs{e}\ \ (\text{mod}\ \abs{e}) ,
\end{align}
respectively. 
They are equal to the disclination charges in Eq.~(\ref{eq:general disclination formula}) with $\omega=b,\ c$.

\bibliography{reference}
\end{document}